\documentclass{elsart}

\usepackage{amssymb}
\usepackage{amsfonts}
\usepackage{amsmath}
\usepackage{graphicx}

\journal{Physics Letters A}

\begin{document}

\begin{frontmatter}

\title{An Architecture of Deterministic Quantum Central Processing Unit}

\author[mphy]{Fei Xue},
\ead{Feixue@mail.ustc.edu.cn}
\author[mphy]{Zeng-Bing Chen}
\author[mphy]{Mingjun Shi}
\author[mphy]{Xianyi Zhou}
\author[mphy]{Jiangfeng Du}
\author[mphy]{Rongdian Han}


\address[mphy]{Department of Modern Physics, University of Science and Technology of China,
Hefei, 230027, People's Republic China}

\date{Apr. 22, 2003}

\begin{abstract}
We present an architecture of QCPU(Quantum Central Processing
Unit) which is based on the discrete quantum gate set. QCPU can be
programmed to approximate any n-qubit computation in a
deterministic fashion. It can be built efficiently to implement
computations with any required accuracy. QCPU makes it possible to
implement universal quantum computation with a fixed, general
purpose hardware.
\end{abstract}


\end{frontmatter}

\section{Introduction}
Quantum information processing offers great advantages both for
quantum communication and quantum computation \cite{cite1,cite2}.
The latter is implemented by unitary operations on a set of
two-level systems known as qubits. These unitary operations are
usually decomposed as quantum gate arrays. Depending on what
unitary operation is desired, different gate arrays are used
\cite{Barenco}. By contrast, a classical computer can be
implemented as a fixed classical gate array--{\it CPU} (Central
Processing Unit), into which is input a {\it program}, and {\it
data}. The program specifies the operations to be performed on the
data. CPU can be programmed to perform any possible function on
the input data. Implementing different operations with different
{\it software} (program) rather than different {\it hardware}
(circuit of gates) is preferable because we could verify whether
it has been prepared correctly or not before using the software,
and we can discard or repair it with little cost if it is found to
be faulty. In contrast, if hardware fails badly during the
execution of a computation, for example some gates in the circuit
are found be set wrong or failed, it might be necessary to build a
whole new hardware.

The possibility to build analogous {\it QCPU} (Quantum Central
Processing Unit) was first studied by Nielsen and Chuang
\cite{Nielsen}. The problem was originally formulated in term of
a programmable array of quantum gates, which can be described as
a fixed unitary operator $G$, that acts on both the program and
the data. The initial state, $P_U$ , of the {\it program
register} stores information about the unitary operation U that
is going to be performed on a {\it data register} initially
prepared in a state $D$. The total dynamics of the programmable
quantum gate array is given by
\begin{equation}\label{concept}
G( \vert P_U \rangle \otimes \vert D \rangle )= \vert
P_U^{\prime} \rangle \otimes U \vert D \rangle.
\end{equation}
Nielsen and Chuang demonstrated that there does not exist
deterministic universal quantum gate array which can be
programmed to perform any unitary operation. Recently Vidal {\it
et al.} \cite{Vidal} and Kim {\it et al.} \cite{Jaehyun}
respectively proposed schemes to store arbitrary one-qubit unitary
operations in quantum states and to retrieve them with the
probability $p=1-\frac{1}{2^m}$, where $m$ is the number of
qubits used to encode the unitary operation on one qubit. Hillery
{\it et al.} presented a probabilistic quantum processor for
qudits on a single qudit of dimension $N$ \cite{Mark}. The above
schemes are alike in storing unitary operations and retrieving
them from program register precisely, but all of them are in a
probabilistic fashion. Thus for these schemes if $N$ gates are
used in total computation the overall success probability is
$p^N$, which tends to fail exponentially with $N$. And it is not
clear how to implement universal quantum computations by a fixed
hardware.

Either model for classical computer or for quantum computer is to
be realized by physical system finally. However real numbers
require infinite information (and therefore infinite energy) for
their representation. Since there appears to be bounds on the
energy of any physical system (the universe included), we can
only approximate real numbers in computers \cite{Lloyd}. So any
model of computer that is realized by physical system only needs
to represent finite accurate things and to perform computation
with finite accuracy, or in other words discrete model of
computer is sufficient.

Nielsen and Chuang's results do not exclude the possibility of
building a gate array which can be programmed to perform an
subset of unitary operations \cite{Nielsen}. In this paper, we
presented an architecture of QCPU--circuit of gate array that can
implement different quantum operations that are discrete on the
$n-$qubit data register according to the IS(Instruction Sequence)
input into the program register in a deterministic fashion. Using
these discrete quantum gates one can approximate any unitary
operation on data register. It is shown that for any given
computation accuracy $\varepsilon$ QCPU can be built efficiently.

\section{Architecture of Deterministic and discrete Quantum Central Processing Unit}

\begin{figure}[bp]
\begin{center}
\includegraphics{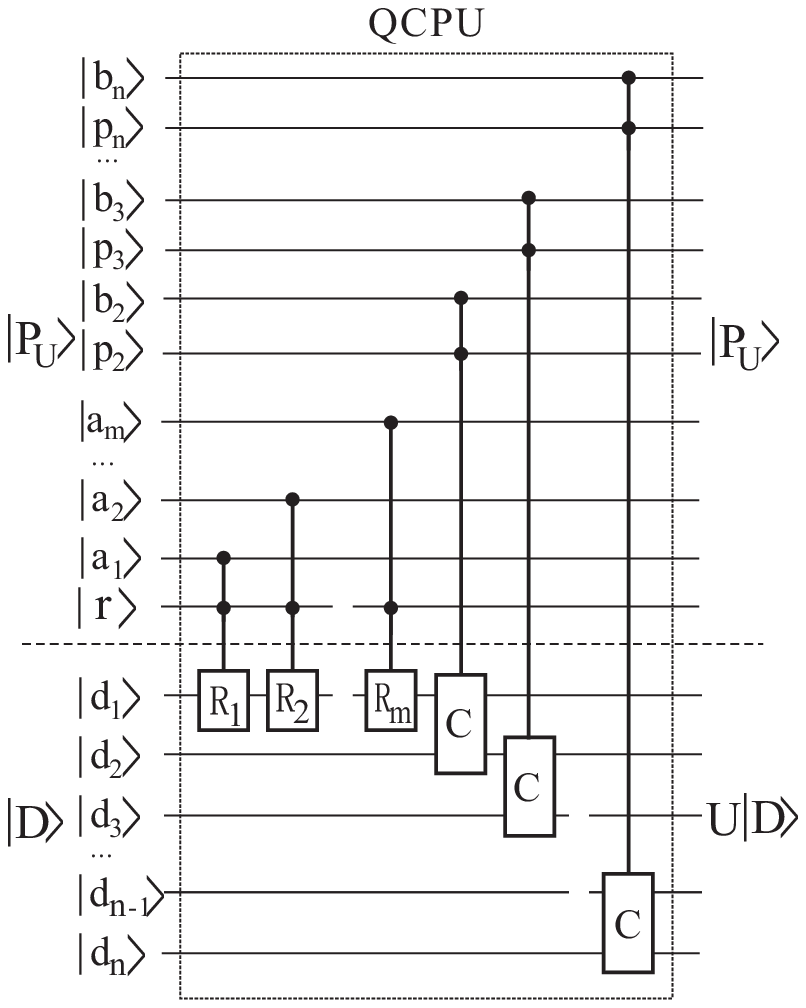}
\end{center}
\caption{\label{qcpu} Architecture of QCPU. The operation
implemented by the gate array in the dot line box corresponds to
the $G$ operation in Eq.(\ref{concept}). $\vert P_U \rangle$ is
the program register,  $\vert D \rangle$ is the data register.}
\end{figure}

\begin{figure}[!]
\begin{center}
\includegraphics{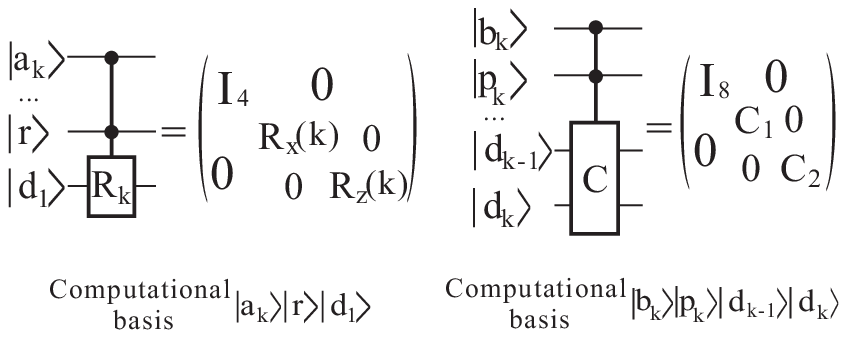}
\end{center}
\caption{\label{element} Functions of element gates in QCPU .}
\end{figure}

The circuit of gate array for QCPU with $n$ qubits in data
register and $1+m+2(n-1)$ qubits in program register is
illustrated in Fig. \ref{qcpu}. The functions of element gates in
QCPU are explained in Fig. \ref{element}, where $I_4$ and $I_8$
are four and eight dimension identity matrix respectively,
$R_x(k)= e^{2^{k-1}\xi i \sigma_x}$, $R_z(k)= e^{2^{k-1}\xi i
\sigma_z}$, $\xi=\frac{2\pi}{2^m}$, and $C_1$ and $C_2$
correspond to CNOT operations,
\begin{eqnarray}
C_1=
\begin{pmatrix}
  1 & 0 & 0 & 0 \\
  0 & 1 & 0 & 0 \\
  0 & 0 & 0 & 1 \\
  0 & 0 & 1 & 0 \\
\end{pmatrix}
, C_2=
\begin{pmatrix}
  1 & 0 & 0 & 0 \\
  0 & 0 & 0 & 1 \\
  0 & 0 & 1 & 0 \\
  0 & 1 & 0 & 0 \\
\end{pmatrix}.
\end{eqnarray}
The controlling qubit $\vert r \rangle$ indicates along which
axis the rotation is implemented. When $\vert r \rangle$ is in
the state $\vert 0 \rangle$ the rotation is along x-axis, and
when $\vert r \rangle$ is in the state $\vert 1 \rangle$ the
rotation is along z-axis. The qubit $\vert a_k \rangle$ indicates
whether the operation is implemented on the qubit $\vert d_1
\rangle$ or not. When $\vert a_k \rangle$ is in the state $\vert
0 \rangle$ the operation is not implemented, and when $\vert a_k
\rangle$ is in the state $\vert 1 \rangle$ the operation is
implemented. The qubit $\vert p_k \rangle$ indicates which qubit
is the control qubit for the CNOT gate on $\vert d_{k-1} \rangle
\vert d_k \rangle$. When $\vert p_k \rangle$ is in the state
$\vert 0 \rangle$ qubit $\vert d_{k-1} \rangle$ is the control
qubit, and when $\vert p_k \rangle$ is in the state $\vert 1
\rangle$ qubit $\vert d_{k} \rangle$ is the control qubit. The
qubit $\vert b_k \rangle$ indicates whether the CNOT gate is
implemented or not. When $\vert b_k \rangle$ is in the state
$\vert 0 \rangle$ the operation is not implemented, and when
$\vert b_k \rangle$ is in the state $\vert 1 \rangle$ the
operation is implemented.

We call the list
\begin{equation}\label{plist}
\vert b_n \rangle \vert p_n \rangle ... \vert b_3 \rangle \vert
p_3 \rangle \vert b_2 \rangle \vert p_2 \rangle \vert a_m \rangle
... \vert a_2 \rangle \vert a_1 \rangle \vert r \rangle
\end{equation}
as {\it Instruction}, and the sequence of Instruction as
IS(Instruction Sequence). $\vert b_n \rangle \vert p_n \rangle
... \vert b_3 \rangle \vert p_3 \rangle \vert b_2 \rangle \vert
p_2 \rangle $ indicates where and which CNOT gate is to be
implemented on the data register. $\vert a_m \rangle ... \vert a_2
\rangle \vert a_1 \rangle$ indicates the angle that is to rotate.
$\vert r \rangle$ indicates which axis the rotation is along. By
inputting corresponding Instruction into the program register,
CNOTs on the data register and rotation along x-axis or z-axis on
$\vert d_1 \rangle$ can be implemented by QCPU in a deterministic
way. Thus any unitary operation on data register can be
approximated by QCPU with the designed IS.

\section{How does the QCPU work}

\begin{figure}[tp]
\begin{center}
\includegraphics{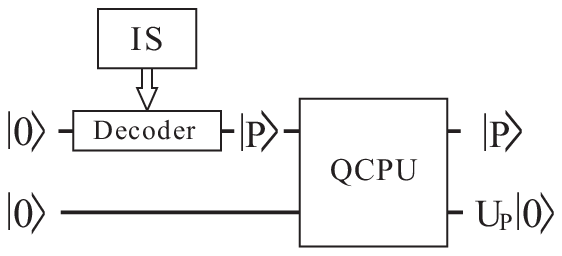}
\end{center}
\caption{\label{circuit1} Working mode one, expanding computation
in the time sequence. The decoder drives the program register to
the special state according to the IS.}
\end{figure}

\begin{figure}[tp]
\begin{center}
\includegraphics{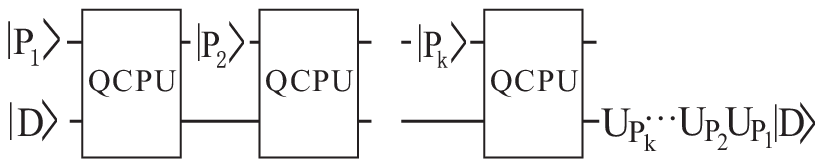}
\end{center}
\caption{\label{circuit2} Working mode two, expanding computation
in the space sequence.}
\end{figure}

Let us explain how the QCPU works. Quantum states are dominated by
quantum laws, such as non-clone principle and measuring collapse.
It is impossible to reset information of a qubit in an unknown
state\cite{Zurek,Kumar}. So it is not straightforward for the
basic operations such as inputting Instructions to program
register. We will describe two working modes of the QCPU, which
are illustrated in Fig. \ref{circuit1} and Fig. \ref{circuit2}.

The first working mode expands computation in the time sequence,
and needs only one QCPU, and no measuring is needed during the
computation. In this working mode we need to initialize program
register to some known state, such as $\vert 00...0 \rangle$, and
to design IS(programs) before implement the computation. Steps of
serial processing of QCPU are illustrated in Table.
\ref{processing}. We use the notation $\vert S \rangle$ to stand
for program register or data register, if $S$ is in capital form,
or to indicate the state of it, if $S$ is in capital form with
subscript of number.

\begin{table}[tp]
\label{processing}\caption{Steps of serial processing of QCPU in
time sequence.}
\begin{tabular}{cccccccc}
\hline
Step & $t_0$ & $t_1$ & $t_2$ & $t_3$ & $t_4$ & $t_5$ & ... \\
\hline\hline \\
$\vert P \rangle$ & $\vert 0 \rangle $ & $\vert P_0 \rangle $ &
$\vert P_0 \rangle $ & $\vert P_1 \rangle $ & $\vert P_1
\rangle $ & $\vert P_2 \rangle $ & ... \\
$\vert D \rangle$ & $\vert 0 \rangle $ & $\vert 0 \rangle $ &
$U_{P_0} \vert 0 \rangle $ & $U_{P_0}\vert 0 \rangle $ &
$U_{P_1}U_{P_0}\vert 0
\rangle $ & $U_{P_1}U_{P_0}\vert 0 \rangle $ & ... \\
\hline \\
Decoder & - & $U_{P_0}$ & - & $U_{P_1}U_{P_0}^{-1}$ & - & $U_{P_2}U_{P_1}^{-1}$ & ... \\
QCPU & - & - & $\rightarrow$ & - & $\rightarrow$ & - & ... \\
\hline
\end{tabular}
\end{table}

To make it more clear on how the operations are encoded in IS,
here we give some examples of IS: specifically we suppose that
data register have 2 qubits, and $m=3(\xi = \frac{\pi}{4})$, then
program register have 6 qubits $ \vert b_2 \rangle \vert p_2
\rangle \vert a_3 \rangle \vert a_2 \rangle \vert a_1 \rangle
\vert r \rangle$.

\begin{eqnarray} \nonumber
operations && Instruction \text{ } or \text{ } IS \\
R_x(\frac{3\pi}{2})&:& 001100 \\
R_z(\frac{5\pi}{4})&:& 001011\\
Swap(d_1, d_2)&:& 100000,110000,100000.
\end{eqnarray}

The angle $\theta$ of rotation is calculated as
\begin{eqnarray} \label{angleor}
\theta=(\alpha_1 2^0 + \alpha_2 2^1 + ... + \alpha _m 2^{m-1})
\frac{2\pi}{2^m},
\end{eqnarray}
where $\alpha_i={0,1}$, $(i=1, 2, ..., m)$.

The second working mode expands computation in the space
sequence, so many QCPUs are needed according to the number of
Instructions used in IS. The advantage of this working mode is
that it can accept the Instruction by teleportation that is
unknown to user.

In both working modes non-unitary operations can be implemented
by performing measurement right after operating QCPU. QCPU can
implement superposed unitary operations by having the program
register in superposition state, and implement entangled unitary
operations by having the program register in entangled state.
Suppose $\vert P \rangle $ is in state $\vert P_0 \rangle + \vert
P_1 \rangle$, then
\begin{eqnarray} \label{entangledo}\nonumber
&&G(( \vert P_0 \rangle + \vert P_1 \rangle ) \otimes \vert D
\rangle ) \\ \nonumber = &&G(\vert P_0 \rangle \otimes \vert D
\rangle + \vert P_1 \rangle \otimes \vert D \rangle) \\ \nonumber
= &&G(\vert P_0 \rangle \otimes \vert D \rangle) + G(\vert P_1
\rangle \otimes \vert D \rangle) \\
= &&\vert P_0 \rangle \otimes U_{P_0}\vert D \rangle + \vert P_1
\rangle \otimes U_{P_1}\vert D \rangle
\end{eqnarray}
One simplest example is that the operation $G$ is the CNOT gate.
When the control-qubit is in superposition such as
$\frac{1}{\sqrt[]2}(\vert 0 \rangle + \vert 1 \rangle)$, the
unitary operations that implemented by the CNOT gate are
superposed: NOT operation and identity operation are performed in
a superposed way on the controlled-qubit. This makes the QCPU very
different from the classical CPU.

If we do not plan to utilize the benefits brought by the
quantumness of the program register, than we may built hybrid
QCPU which has the data register in qubits and the program
register in classical bits. This hybrid QCPU is also capable of
implementing any quantum computation on n-qubit data register, but
only need $n$ qubits and $1+m+2(n-1)$ classical bits.

\section{The QCPU Can Be Built Efficiently To Approximate Arbitrary Unitary Operations To Any Given Precision }

Now let us explain why arbitrary unitary operation on the data
register can be approximated by QCPU. It was explained above that
QCPU can implement CNOTs on the data register, and can
approximate rotations on $ \vert d_1 \rangle $ along x-axis and
z-axis with accuracy $\xi= \frac{2\pi}{2^m}$. Then based on the
fact that any single qubit operation can be expressed as at most
three rotations about two non-parallel axes, QCPU is capable of
approximating arbitrary single qubit operation on $ \vert d_1
\rangle $. With the help of the swap gate which can be
accomplished by three CNOTs, QCPU can approximate any single
qubit operation on any qubit of the data register. We have known
that CNOTs and arbitrary single qubit operations together can
complete universal quantum computation \cite{Barenco}. Therefore
QCPU proposed above can approximate arbitrary unitary operation on
the data register.

The error caused by approximate rotation rather than precision
rotation is estimated below. Angles denoted by Eq.(\ref{angleor})
symmetrically distribute between [$0,2\pi$], and are implemented
by QCPU precisely. So the error that QCPU approximates an
arbitrary angle is less than $\frac{2\pi}{2^{m}}$. All
computation gives outputs at last. The approximation of the
unitary transformations will lead to errors in the resultant
output states. The errors of the outputs, which are from the
errors of $\theta$, are calculated below.

Rotation along $a$-axis can be expressed as $U_a( \theta )=e^{i
\theta \sigma_a}$, where $\sigma_a$ is a Pauli matrix, $a=x,y,z$.
Then
\[ \delta U_a( \theta )= i \sigma_a e^{i \theta \sigma_a}
\delta \theta . \] The variation of the output state $U_a( \theta
)\vert \psi \rangle$ reads
\[ \delta (U_a( \theta )\vert \psi \rangle)= \delta U_a(
\theta ) \vert \psi \rangle. \] The variation of the probability
$\vert \langle \varphi \vert U_a( \theta )\vert \psi \rangle
\vert ^2$ of projecting $U_a( \theta )\vert \psi \rangle$ onto a
state $\vert \varphi \rangle$ reads
\begin{eqnarray}\label{Oonstate}\nonumber
\delta P=\langle \varphi \vert \delta U_a( \theta ) \vert \psi
\rangle \langle \psi \vert \ U_a^\dagger( \theta )
\vert \varphi \rangle\\
\nonumber + \langle \varphi \vert U_a( \theta ) \vert \psi
\rangle \langle \psi \vert \delta U_a^\dagger( \theta ) \vert
\varphi \rangle ,
\end{eqnarray}
So we have $\vert \delta P \vert \le  2 \vert \delta \theta
\vert$.

It is shown that when the output state $U_a( \theta )\vert \psi
\rangle$ are projected on the state $\vert \varphi \rangle$, the
error caused the approximate rotation is linear with the accuracy
of $\theta$. When there are $N$ such unitary transformations $U=
U(\theta_1)U(\theta_2)...U(\theta_N)$ before the measurement, we
have
\begin{eqnarray}
\vert \delta P \vert \le 2 ( \vert \delta \theta_1 \vert +
\vert\delta \theta_2 \vert + ... + \vert \delta \theta_N \vert )
\end{eqnarray}
Thus in order to approximate a computation that is implemented by
$N$ approximate operations to an overall accuracy $\varepsilon$,
each operation only needs to be accurate to $\varepsilon/N$
\cite{Ethan}.

The number of qubits, which is needed in the program register to
approximate any computation on n-qubit data register to accuracy
$\varepsilon$, is calculated below. We know that any $2^n \times
2^n$ unitary operation can be implemented by $O(n^34^n)$ CNOTs and
single qubit operations \cite{Barenco}. Each CNOT or single qubit
operation can be implement by at most $3n$ Instructions in QCPU.
So the number of Instructions needed to implement arbitrary
operation on n qubits is $O(n^44^n)$. If the overall computation
precision is $\varepsilon$ then each Instruction need to be
accurate to $o(\varepsilon/(n^44^n))$. Rotation implemented by
single Instruction in QCPU is accurate to about
$\frac{1}{2^{m}}$. So
\begin{eqnarray} \nonumber
m=&& O(\log_2 \frac{n^44^n}{\varepsilon}) \\
= && O(4\log_2n + 2n -\ln\varepsilon )
\end{eqnarray}
The number of qubits needed in program register is $1+O(4\log_2n
+ 2n -\ln\varepsilon)+2(n-1)=O(n)-\ln\varepsilon$. It is linear
with the number of data qubits and $\ln\varepsilon$, therefore
QCPU can be built efficiently to approximate universal quantum
computation on any number of qubits to any given computation
accuracy $\varepsilon$.

{\it QCPU} enables one to realize quantum computations with
different {\it software} rather than different {\it
hardware}(circuit of gates). A quantum software is a sequence of
particular programmed IS that makes a QCPU to perform a specific
task. The notation of {\it quantum software} was first used by
Preskill \cite{Preskill}. He used it in a little different way.
But the virtue of quantum software itself is alike. Realizing
quantum computation with software rather than hardware is
preferable for a lot of reasons. If the IS are in quantum states,
QCPU in working mode two is capable of implementing quantum remote
control introduced by Huelga {\it et al.} in
\cite{Huelga1,Huelga2}. One important virtue of quantum software
will be to ensure that quantum computers function reliably.
Because quantum states are very fragile, the quantum-computing
hardware needs to meet very demanding specifications. Quantum
computer can achieve acceptable reliability by applying principles
of quantum-error correction \cite{Steane1,Steane2}. Quantum-error
correction schemes would be most conveniently implemented with
quantum software. Some time in the future the fault-tolerant
quantum computer, which could possibly be dominated by quantum
software, may achieve processing speeds far surpassing the
classical computers.

\section{Conclusion}

To summarize, we have presented an architecture of QCPU which
operates on a $n$-qubit data register and is capable of completing
any unitary operation with accuracy $\varepsilon$ in a
deterministic way. It contains $O(n)-\ln\varepsilon$ three-qubit
gates and $(n-1)$ four-qubit gates and needs $O(n)-\ln\varepsilon$
qubits as its input. Therefore it can be built efficiently. Each
gate only concerns at most four qubits, this may be appreciated in
real implementations of the quantum computer. We have described
two working modes of QCPU. QCPU have the ability to implement
superposed and entangled unitary operations on the data register,
which is shown by Eq.(\ref{entangledo}). This ability could help
us to implement more efficient algorithm--the number of IS needed
to implemented a algorithm might be much smaller than the upper
bounds $O(n^44^n)$. One noteworthy quality of our architecture is
that it can approximate n-qubit universal quantum computation with
only n qubit plus $O(n)-\ln\varepsilon$ classical bits as its
input. So with sufficiently sophisticated hardware QCPU could run
on classical software but implement the quantum computation.

QCPU made it possible to put the solution of the problem into
software rather than hardware. This is an important step to the
general quantum computer because it is impracticable to build a
special hardware for each problem. In the hardware of quantum
computer qubits must preserve coherence during operations. Thus
the scale(the number of qubits in quantum hardware) and the the
operating time of quantum computer have a lot of limitations. When
the space complexity(scale) or time complexity(time) is beyond the
capability of the hardware, programming technique in software may
help us to make the balance of space complexity and time
complexity. This possibly enlarges our computation ability under
specifical hardware technique. Then like the art of programming in
classical computer, the art of programming in quantum computer
will settle various kinds of problem with a fixed and general
purpose hardware, and the art of programming in quantum computer
will also improve the efficiency of algorithms by just refining
the program(IS).

We thank Prof. Y.D. Zhang for useful comments. This work was
supported by the National Nature Science Foundation of China
(Grants No. 10075041, No. 10075044 and No. 10104014), and the
National Fundamental Research Program(Grant No. 2001CB309300).

\end{document}